\documentclass[reprint,superscriptaddress,showkeys,amsmath,amssymb,aps,twocolumn,longbibliography]{revtex4-2} % chktex 8
\setcitestyle{super,open={},close={}}

\usepackage{orcidlink}
\usepackage{charter}
\usepackage{mathptmx}
\usepackage{bm}
\usepackage{microtype}
\usepackage[separate-uncertainty=true, multi-part-units=single]{siunitx}
\usepackage[version=4]{mhchem}
\renewcommand{\vec}[1]{\boldsymbol{#1}}
\usepackage{graphicx}
\graphicspath{{figs/}}

\usepackage{booktabs}
\usepackage[dvipsnames, svgnames, table, xcdraw]{xcolor}
\usepackage{hyperref}
\hypersetup{colorlinks, breaklinks, linkcolor=black, urlcolor=black, citecolor=black, anchorcolor=black} % new
\usepackage{fancyhdr}
\usepackage{titlesec}
\titleformat{\section}[runin]{\normalfont\bfseries\itshape}{\thesection}{0.5em}{}[\,\,\,--] % chktex 6
\titleformat{\subsection}[runin]{\normalfont\itshape}{\thesubsection}{0.5em}{}[\,\,\,--] % chktex 6
\titlespacing*{\section}{0pt}{*1}{5pt}
\titlespacing*{\subsection}{0pt}{*1}{5pt}

\begin{document}
\title{Realizing microrheological response of configurable viscoelastic media with a dynamic optical trap}
\author{Sanatan Halder \orcidlink{0009-0002-2457-0449}} % chktex 8
\email{sanatanh@iitk.ac.in}
\affiliation{Department of Physics, Indian Institute of Technology Kanpur, Kanpur - 208016, India} % chktex 8
\author{Manas Khan \orcidlink{0000-0001-6446-3205}} % chktex 8
\email{mkhan@iitk.ac.in}
\affiliation{Department of Physics, Indian Institute of Technology Kanpur, Kanpur - 208016, India} % chktex 8
\begin{abstract} \bfseries
  \noindent The local viscoelastic (VE) environment governs the motion of an embedded microsphere and consequently, pertinent dynamical phenomena. 
  However, studying such phenomena with varying VE properties remains challenging for various reasons, including the strong coupling among the VE parameters and their dependence on experimental conditions, such as temperature. 
  Here, we demonstrate the experimental realization of configurable VE media with broad variations, wherein the VE properties can be systematically and independently tuned, employing a dynamic optical trap.
  Specifically, the dynamics of a particle in a slowly diffusing optical trap provides the linear microrheological response of single-relaxation VE fluids, namely Jeffreys or Maxwell-Voigt (MV) fluids, where the trap strength and its diffusion coefficient regulate the elastic response and the low-frequency viscosity, respectively.
  The characteristic features in the mean square displacement (MSD) of the trapped particle match those of a probe particle in real MV fluids, and the simulation results following harmonically bound Brownian particle with long-time diffusion model describing single-relaxation complex fluids.
  Our scheme is further validated by demonstrating excellent quantitative agreement between the experimentally observed MSDs of the trapped bead and those from the corresponding analytical predictions.
  We extend the applicability of this scheme for realizing the microrheological response of double-relaxation VE media by incorporating appropriately correlated noise in the trap trajectory, signifying its validity for any linear VE media with multiple relaxations. Our scheme can be further extended to realize probe particle dynamics in an active VE environment, e.g., an entangled network of active polymers, by translating the trap along an active Brownian trajectory. 
  Therefore, our scheme enables systematic microrheological studies in VE regimes that are otherwise challenging to realize or not readily accessible with real materials.
\end{abstract}
\maketitle

\section*{Introduction}
Complex fluids, including wormlike micelles~\cite{shikataNonlinearViscoelasticBehavior1988, catesStaticsDynamicsWormlike1990, kernRheologicalPropertiesSemidilute1992, bellourBrownianMotionParticles2002, cardinauxMicrorheologyGiantmicelleSolutions2002, hassanMicrorheologyWormlikeMicellar2005}, emulsions~\cite{masonDiffusingwavespectroscopyMeasurementsViscoelasticity1997}, entangled polymer solutions~\cite{ferryViscoelasticPropertiesPolymers1980, catesReptationLivingPolymers1987, masonParticleTrackingMicrorheology1997, dasguptaMicrorheologyPolyethyleneOxide2002, chengRotationalDiffusionMicrorheology2003, vanzantenBrownianMotionColloidal2004, andablo-reyesMicrorheologyRotationalDiffusion2005}, biopolymer networks~\cite{gardelMicrorheologyEntangledFActin2003, wilkingOpticallyDrivenNonlinear2008, shabaniverkiCharacterizingGelatinHydrogel2017}, and active polymers~\cite{mizunoNonequilibriumMechanicsActive2007, eisensteckenInternalDynamicsSemiflexible2017, tejedorReptationActiveEntangled2019, winklerPhysicsActivePolymers2020, tejedorMolecularDynamicsSimulations2023}, exhibit viscoelastic (VE) properties that govern microscale transport and relaxation dynamics~\cite{squiresFluidMechanicsMicrorheology2010}.
The frequency-dependent shear moduli of these materials profoundly influence the motion of embedded colloidal probes~\cite{masonRheologyComplexFluids1996}.
This connection between probe dynamics and material properties offers a direct window into the VE response of complex fluids.
The resulting framework is used in microrheology, where the translational and orientational thermal motion of probe particles embedded in complex fluids reveals the VE properties of their environment through the generalized Stokes-Einstein relation (GSER)~\cite{masonOpticalMeasurementsFrequencyDependent1995, masonRheologyComplexFluids1996, masonParticleTrackingMicrorheology1997, masonDiffusingwavespectroscopyMeasurementsViscoelasticity1997, gittesMicroscopicViscoelasticityShear1997, masonEstimatingViscoelasticModuli2000, crockerTwoPointMicrorheologyInhomogeneous2000, chengRotationalDiffusionMicrorheology2003, andablo-reyesMicrorheologyRotationalDiffusion2005, squiresFluidMechanicsMicrorheology2010, furstMicrorheology2020}. 
Constitutive models representing linear VE responses greatly aid in the understanding and interpretation of probe dynamics.
Many studies employ the Jeffreys fluid model, which describes single-relaxation complex fluids with a Voigt element, constituted by a spring and dashpot in parallel, and another dashpot in series~\cite{raikherTheoryBrownianMotion2010, raikherBrownianMotionViscoelastic2013}. 
The Voigt element is the simplest rheological model for an elastoviscous medium, capturing elastic recovery alongside viscous dissipation.
The Maxwell-Voigt (MV) fluid extends this framework by connecting a Maxwell element with a Voigt element in series, where the Maxwell element is represented by a spring and dashpot in series, with a shared spring constant $k$, providing an enhanced description of materials with pronounced elastic plateaus~\cite{vanzantenBrownianMotionSingle2000, wilhelmRotationalMagneticEndosome2003, wilhelmMagneticNanoparticlesInternal2009, grimmBrownianMotionMaxwell2011, rusakovBrownianMotionFluids2015}.

Knowledge of the local microrheological response of the medium is crucial for comprehending complex transport phenomena across mesoscopic length scales, ranging from intracellular transport to microorganism motility in the mucus~\cite{lauMicrorheologyStressFluctuations2003, celliHelicobacterPyloriMoves2009, pattesonRunningTumblingColi2015, gomez-solanoDynamicsSelfPropelledJanus2016}. 
However, studying probe dynamics in real complex fluids in the desired VE regimes or in specific settings presents significant experimental challenges.
The VE properties of these systems depend sensitively on temperature, concentration, and sample age~\cite{ferryViscoelasticPropertiesPolymers1980, doiTheoryPolymerDynamics2013, catesStaticsDynamicsWormlike1990}, whereas the microrheological response varies with active forcing~\cite{squiresNonlinearMicrorheologyBulk2008}, particle size~\cite{luProbeSizeEffects2002, vanzantenBrownianMotionColloidal2004}, and surface chemistry~\cite{mcgrathMechanicsFActinMicroenvironments2000}.
Moreover, material parameters such as relaxation time, viscosity, and elastic modulus are inherently coupled, making it impossible to vary a single parameter independently for systematic studies.
These limitations underscore the necessity of developing experimental approaches in which the VE environment can be precisely configured and systematically tuned.

\begin{figure}[b]
  \centering
  \includegraphics[width=85mm]{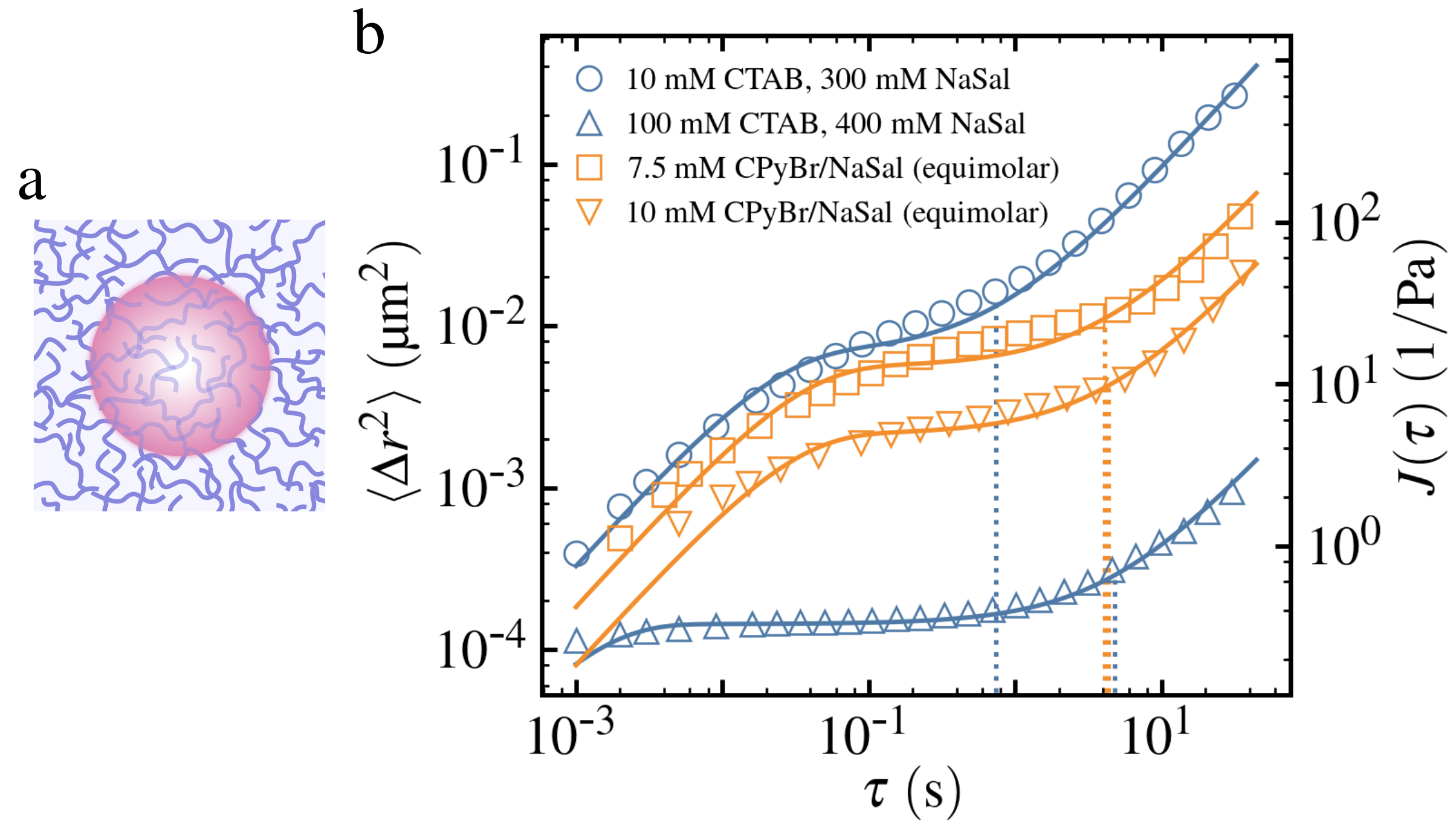}%
  \caption{Typical microrheological response of single-relaxation viscoelastic (VE) fluids. 
    (a) The schematic shows a probe microsphere embedded in a complex fluid represented by entangled wormlike micelles. 
    (b) Microrheological responses of cetyltrimethylammonium bromide--sodium salicylate (\ce{CTAB-NaSal}) and cetylpyridinium bromide--sodium salicylate (\ce{CPyBr-NaSal}) are shown at various concentrations, characterized by the MSD ($\left\langle \Delta r^2 \right\rangle$, left axis) of an embedded \SI{1.98}{\micro\meter} polystyrene (PS) probe particle and derived creep compliance ($J$, right axis) using Eq.~\ref{eq:gser}. 
    Open symbols represent the experimentally captured data, and the solid lines of the same color are fits to Eq.~\ref{eq:msd-pbp}. 
    The vertical dotted lines indicate the relaxation time $\lambda$.}%
  \label{fig:msd-real}
\end{figure}

\begin{figure*}[t]
  \centering
  \includegraphics[width=175mm]{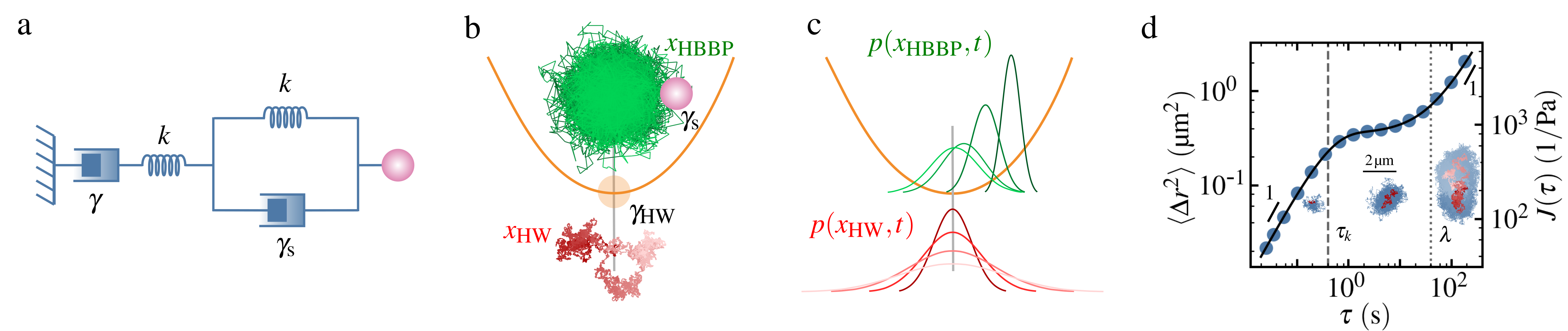}%
  \caption{Harmonically bound Brownian particle (HBBP) with long-time diffusion model describing the microrheological response of an MV fluid.
    (a) A schematic shows the spring-dashpot model of a Maxwell-Voigt (MV) fluid, constituted by two springs with a common spring constant ($k=6\pi a G_\mathrm{p}$, with probe radius $a$ and plateau modulus $G_\mathrm{p}$) and dashpots with different damping coefficients ($\gamma_\mathrm{s}$ and $\gamma$) corresponding to the high- and low-frequency viscosities ($\eta_{\mathrm{s}}$ and $\eta$), respectively.
    (b) Pictorial illustration of the HBBP with long-time diffusion model exhibits the typical confined dynamics of a HBBP (magenta filled circle, friction coefficient $\gamma_\mathrm{s}$) with instantaneous position $x_\mathrm{HBBP} (t)$ (green), and the long-time diffusion of the harmonic well (HW, friction coefficient $\gamma_\mathrm{HW}$) by a representative trajectory $x_\mathrm{HW} (t)$ (red).       
    (c) Time evolution of the position distribution of the HBBP in reference to the HW, $p (x_{\mathrm{HBBP}}, t)$ (green), and that of the HW in the laboratory frame, $p (x_{\mathrm{HW}}, t)$ (red), are shown at four progressing times (lighter shades indicate later times). 
    (d) Filled symbols show the simulated MSD ($\left\langle \Delta r^2 \right\rangle$, left axis) of a $1.98~\unit{\micro\meter}$ probe sphere in an MV fluid from Langevin dynamics (Eq.~\ref{eq:le-pbp-mv}) and the corresponding creep compliance ($J$, right axis), with black line showing fits to the theoretical prediction of the model (Eq.~\ref{eq:msd-pbp}).
        The dashed and dotted vertical lines mark the timescales $\tau_k$ and $\lambda$. 
        The blue trajectories represent the probe dynamics at three different times ($t = \{1, 10, 100\}\;\unit{\second}$), with the slowly diffusing HW overlaid in red.}%
  \label{fig:model-mv}
\end{figure*}

Here, we demonstrate an experimental scheme using a dynamic optical trap to obtain the linear microrheological response of a configurable VE medium.
Drawing motivation from the harmonically bound Brownian particle (HBBP) with long-time diffusion model~\cite{khanRandomWalksColloidal2014}, we steer an optical trap, which harmonically confines a Brownian probe particle, along an appropriate trajectory in a Newtonian fluid, constructing an emergent VE environment with a desired microrheological response. Notably, the VE response is solely realized by the dynamic optical trap without the presence of any real elastic network.
This approach enables us to fully configure the relaxation dynamics of the medium, manifesting single-relaxation, double-relaxation, or even active relaxation, which is observed in far-from-equilibrium entangled networks of active polymers.
While moving the optical trap along a free diffusive trajectory generates a single relaxation, as observed in the MV or Jeffrey fluid, trap trajectories with correlated translational noise give rise to another elastic confinement regime at a longer time, manifested as a second plateau, with an eventual second relaxation. 
Active relaxation is observed when the trap is moved following the trajectory of an active Brownian particle. 
Furthermore, this scheme provides independent tunability of various VE parameters, such as the high-frequency viscosity ($\eta_{\mathrm{s}}$), plateau modulus ($G_{\mathrm{p}}$), and relaxation time ($\lambda$), using an appropriate solvent with viscosity $\eta_{\mathrm{s}}$, setting optical trap stiffness $k = 6\pi a G_{\mathrm{p}}$, and regulating the diffusion profile of the optical trap, respectively, where $a$ is the Stokes radius of the probe sphere.
We validate our scheme by comparing the microrheological response obtained from the probe dynamics in the diffusing optical trap with that of real single-relaxation VE fluids and the theoretical predictions of the HBBP with long-time diffusion model.

\section*{Single-relaxation VE fluids}\label{sec:single-relax}
We measured the microrheological response of a polystyrene (PS) probe particle ($2a = \SI{1.98}{\micro\meter}$) by recording and analyzing its dynamics in two types of entangled wormlike micellar solutions: cetyltrimethylammonium bromide--sodium salicylate (CTAB--NaSal) and cetylpyridinium bromide--sodium salicylate (CPyBr--NaSal),  across various concentrations of surfactants and counterions (Video~\href{https://drive.google.com/file/d/1_Gz15QQK7ICtZgLIAA-o8UdsBIs4g8Lu/view?usp=share_link}{S1}).
In all cases, the mean square displacement (MSD) of the probe reveals three distinct regimes (Fig.~\ref{fig:msd-real}). 
At short time-lags, the particle undergoes free diffusion, manifesting the high-frequency solvent viscosity $\eta_{\mathrm{s}}$. 
At intermediate time-lags, the MSD reaches a plateau, signifying the elastic confinement imposed by the entangled micellar network.
At long time-lags, i.e., $\tau > \lambda (= \eta / G_{\mathrm{p}})$, the plateau gives way to a diffusive regime corresponding to the slow Maxwell relaxation, where $\lambda$ and $\eta$ are the relaxation time and low-frequency viscosity, respectively.
These three regimes, namely, short-time free diffusion, an intermediate elastic plateau, and long-time diffusion, constitute the hallmarks of single-relaxation VE fluids.
Materials exhibiting this behavior with a pronounced elastic plateau are well described by the MV fluid model, in which the Voigt and Maxwell elements are connected in series and share a common spring constant $k$~(Fig.~\ref{fig:model-mv}a).
We fitted the measured MSDs to the theoretical prediction of the HBBP with long-time diffusion model~(Fig.~\ref{fig:model-mv}b), and their excellent agreement validated the MV description of the VE fluids across all concentrations examined.

\section*{HBBP with long-time diffusion model}
The dynamics of a probe particle in an MV fluid can be modeled by that of an HBBP with long-time diffusion (Fig.~\ref{fig:model-mv}b)~\cite{khanRandomWalksColloidal2014}.
In this model, the probe particle of radius $a$ diffuses in a solvent of viscosity $\eta_{\mathrm{s}}$ under harmonic confinement with a spring constant $k = 6\pi a G_{\mathrm{p}}$, capturing the Voigt-solid response, where the plateau modulus $G_{\mathrm{p}}$ is defined by the trap stiffness $k$.
The center of this harmonic well (HW) itself undergoes slow Brownian diffusion, incorporating the long-time Maxwell relaxation, which can be regulated by a tunable friction coefficient $\gamma_{\mathrm{HW}} = 6\pi a \eta_{\mathrm{HW}} = 6\pi a \eta$, where $\eta_{\mathrm{HW}}$ is the viscosity associated with the diffusion of the HW, and $\eta$ is the low-frequency viscosity of the MV fluid..
The overdamped Langevin equations representing the probe dynamics consist of two parts: one for the HBBP, describing the position of the probe $x_{\mathrm{HBBP}}$ relative to the HW center, and the other for the diffusion of the HW center $x_{\mathrm{HW}}$. The equations are given by
\begin{subequations}
  \begin{align}
    \dot{x}_{\mathrm{HBBP}}(t) & =-\frac{x_{\mathrm{HBBP}}(t)}{\tau_k} + v_{r, \mathrm{HBBP}}(t) \label{eq:le-pbp-hw}, \quad \text{and}\\
    \dot{x}_{\mathrm{HW}}(t) & = v_{r,\mathrm{HW}}(t), \label{eq:le-free-pbp} 
  \end{align}%
  \label{eq:le-pbp-mv}%
\end{subequations}
\noindent where $\tau_k = \gamma_{\mathrm{s}}/k$ is the equilibration time of the particle, with the drag coefficient $\gamma_{\mathrm{s}} = 6\pi a \eta_{\mathrm{s}}$.
The terms $v_{r,\mathrm{HBBP}}(t)$ and $v_{r, \mathrm{HW}}(t)$ are independent white Gaussian velocity noises with zero mean and correlation $\langle v_{r,i}(t_1)v_{r,j}(t_2) \rangle = 2D_i\delta_{ij}\delta(t_1 - t_2)$, where $D_i$ is the translational diffusivity, and $i,j \in \{\mathrm{HBBP}, \mathrm{HW}\}$.
Hence, the probe dynamics in a Voigt solid, given by $x_{\mathrm{HBBP}} (t)$ (Eq.~\ref{eq:le-pbp-hw}), and Maxwell relaxation, incorporated by $x_{\mathrm{HW}} (t)$ (Eq.~\ref{eq:le-free-pbp}), satisfy the fluctuation-dissipation theorem (FDT)~\cite{rusakovBrownianMotionFluids2015, raikherTheoryBrownianMotion2010, raikherBrownianMotionViscoelastic2013}. Consequently, the resultant motion of the probe particle in an MV fluid, as described by the HBBP with the long-time diffusion model and experimentally realized with a dynamic optical trap, is consistent with the FDT.

The PDF for the position of HBBP at time $t$, denoted by $p(x_{\mathrm{HBBP}}, t)$, is described by the Green's function $G(x_{\mathrm{HBBP}}, t; x_{{\mathrm{HBBP}},0})$~\cite{chandrasekharStochasticProblemsPhysics1943, uhlenbeckTheoryBrownianMotion1930},
\begin{equation}
  G(x_{\mathrm{HBBP}}, t ; x_{{\mathrm{HBBP}},0}) = {[2{\pi}B(t)]}^{-1/2} e^{-{\left(x_{\mathrm{HBBP}}-A(t)\right)}^2 / 2B(t)},
  \label{eq:pdf-greens-func-voigt}
\end{equation}
\noindent where $A(t) = x_{\mathrm{HBBP},0}\; e^{-t/\tau_k}$ and $B(t) = [k_{\mathrm{B}}T/k] \left( 1-e^{-2t/\tau_k}\right)$.
The PDF of $x_{\mathrm{HW}}$ is also Gaussian with time-dependent variance $2D_{\mathrm{HW}} t$, given by
\begin{equation}
  p(x_{\mathrm{HW}}, t) = {\left[2\pi D_{\mathrm{HW}} t \right]}^{-1/2} e^{-{\left(x_{\mathrm{HW}} - x_{{\mathrm{HW}},0}\right)}^2 / 2D_{\mathrm{HW}} t}.
  \label{eq:pdf-maxwell}
\end{equation}
The time evolution of these two PDFs, $p(x_{\mathrm{HBBP}}, t)$ and $p(x_{\mathrm{HW}}, t)$, is shown in ~Fig.~\ref{fig:model-mv}c.

Because the Maxwell and Voigt elements are connected in series, the particle position in the MV fluid is the sum of the two uncorrelated contributions: $x(t) = x_{\mathrm{HBBP}}(t) + x_{\mathrm{HW}}(t)$.
Hence, the MSD of a probe particle in an MV fluid can be derived by adding the MSD of the HBBP and that of the HW, 
\begin{align}
  \langle \Delta r^2(\tau)\rangle=  & 4D_{\mathrm{HBBP}}\tau_k \left(1 - e^{-\tau/\tau_k}\right) + 4D_{\mathrm{HW}}\tau.
  \label{eq:msd-pbp}
\end{align}
Although this MSD expression is derived following the HBBP with long-time diffusion model and is hence given in terms of the model parameters that we directly implement in the experimental realization, it is the standard form describing the probe dynamics in an MV or Jeffreys fluid~\cite{raikherTheoryBrownianMotion2010}. A typical MSD of a probe particle in an MV fluid, modeled as an HBBP with long-time diffusion, is shown in Fig.~\ref{fig:model-mv}d.

The VE properties of the MV fluid can be extracted from the MSD via GSER,
$\tilde{G_{\mathrm{p}}}(s) = k_{\mathrm{B}}T/\pi a s \langle \Delta \tilde{r}^2 (s)\rangle$ where $\left\langle \Delta \tilde{r}^2 (s)\right\rangle = \mathcal{L}\left\lbrace  \Delta r^2 (\tau) \right\rbrace$~\cite{squiresFluidMechanicsMicrorheology2010}.
Thus, the creep compliance $J(\tau)$ is given by
\begin{equation}
  J(\tau) = \frac{\pi a}{k_{\mathrm{B}}T} \langle \Delta r^2 (\tau) \rangle.
  \label{eq:gser}
\end{equation}
This model captures the hallmark features of the microrheological response of single-relaxation VE fluids with a prominent elastic plateau, as observed in entangled wormlike micellar solutions (Fig.~\ref{fig:msd-real}).

\section*{Extension of the model beyond single-relaxation}
The diffusing trap framework also extends to generalized linear viscoelastic (GLVE) materials with multiple relaxation modes~\cite{khanTrajectoriesProbeSpheres2014}. 
Each additional noise correlation in the HW dynamics introduces a new elastic plateau at the correlation time, thereby enabling systematic construction of generalized Maxwell models (GMM) through controlled noise superposition.
We demonstrate this by introducing a second harmonic component with timescale $\tau_{k,2}$ in the HW dynamics, creating an effective double-relaxation VE response.
The resulting particle MSD is given by
\begin{align}
  \langle \Delta r^2(\tau)\rangle=  \sum^2_{i=1} 4D_i\tau_{k, i} \left(1 - e^{-\tau/\tau_{k,i}}\right).
  \label{eq:msd-pbp-double}
\end{align}
The MSD exhibits two distinct plateaus at times $\tau_{k,1}$ and $\tau_{k,2}$, with the first mode relaxing at finite time $\lambda_1$ and the second at $\lambda_2 \to \infty$, corresponding to an HBBP whose bounding potential diffuses within a weaker static HW~\cite{khanTrajectoriesProbeSpheres2014}.

Active VE environments, such as active entangled polymer networks exhibit persistent motion that fundamentally differs from thermal fluctuations~\cite{eisensteckenInternalDynamicsSemiflexible2017, winklerPhysicsActivePolymers2020, tejedorMolecularDynamicsSimulations2023}.
By imposing persistent noise on the HW trajectory, we create an effective active VE environment that captures key features of activity-driven systems.
In this case, we translate the HW following an active Brownian particle (ABP) trajectory, given by
\begin{align}
  \dot{\vec{r}}_{\mathrm{HW}}(t) = V\hat{\phi} + \vec{v}_{r, \mathrm{HW}}(t)
  \label{eq:le-free-abp}%
\end{align}
\noindent where $V$ is the propulsion speed and $\hat{\phi}$ is the unit vector along the propulsion direction, which is governed by the rotational diffusivity $D_{\mathrm{R}}$. The strength of activity is measured by the P\'{e}clet number, $\mathrm{Pe} = V / \sqrt{D_{\mathrm{R}} D_\mathrm{HW}}$, with $\mathrm{Pe} \gg 1$ indicating that active motion dominates thermal diffusion.
This generates persistent motion of the probe particle with a correlation time $\tau_{\mathrm{R}}=1/D_{\mathrm{R}}$.

The MSD of a probe sphere in this active VE medium is obtained by replacing the slow diffusion of the HW in Eq.~\ref{eq:msd-pbp} with the active HW dynamics. Since the translational and rotational dynamics of an ABP are independent, the contribution of the active dynamics to the MSD consists of a translational diffusion part $4D_{\mathrm{HW}}\tau$, identical to the passive case, and an additional propulsion term that loses persistence of direction at timescale $\tau_{\mathrm{R}}$~\cite{tenhagenNonGaussianBehaviourSelfpropelled2009}.
Therefore, the resulting MSD of the probe particle in this active VE environment is given by,
\begin{align}
  \langle \Delta r^2(\tau)\rangle = {}& 4D_{\mathrm{HBBP}}\tau_{k}\left(1 - e^{-\tau/\tau_k}\right) \nonumber\\
                                    & + 4D_{\mathrm{HW}}\tau + 2V^2\tau^2_{\mathrm{R}}\left(\tau/\tau_{\mathrm{R}} + e^{-\tau/\tau_{\mathrm{R}}} - 1\right).
  \label{eq:msd-pbp-active}
\end{align}
The particle exhibits free diffusion at short time-lags $\tau \ll \tau_k$, manifesting the solvent viscosity.
Subsequently, it experiences elastic confinement by the entangled network at $\tau > \tau_k$.
In the post-relaxation regime ($\tau > \lambda$), the particle exhibits super-diffusive motion for $\tau < \tau_R$, transitioning to diffusive dynamics at $\tau > \tau_R$.
This framework enables systematic investigation of passive probe dynamics in active VE environments across varying activity and viscoelasticity.

Notably, the GSER (Eq.~\ref{eq:gser}) applies only to passive VE media, where the FDT holds for single-probe particle dynamics, but not in this case of active relaxation.

\begin{figure*}[t]
  \centering
  \includegraphics[width=165mm]{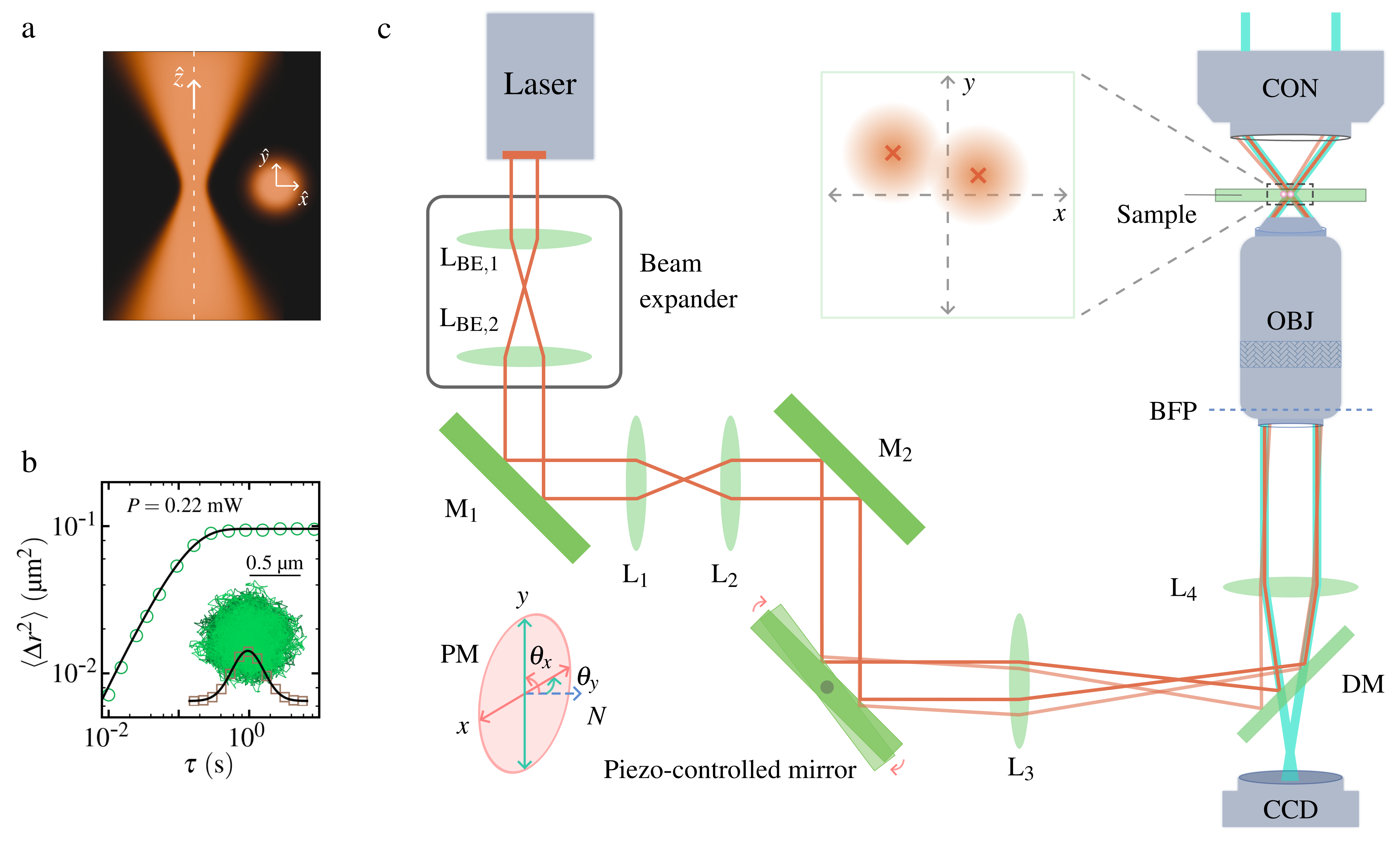}%
  \caption{Experimental realization of the microrheological response of a configurable VE fluid with a dynamic optical trap.
    (a) The orange gradients show the axial and radial intensity profiles of a tightly focused laser beam propagating along $\hat{z}$, creating harmonic confinement at the focal point in the $x$-$y$ plane.
    (b) The MSD ($\langle \Delta r^2 \rangle$) of a \SI{1.98}{\micro\meter} PS probe sphere in a static optical trap exhibits the microrheological response of a Voigt solid, with the corresponding bound trajectory and overlaid long-time position distribution $p(x_{\mathrm{HBBP}})$ shown in the inset.
    (c) The ray diagram shows a steerable optical trap.
        L and M denote the lenses and mirrors in the laser beam path (orange lines). 
        The trap is steered in the focal ($x$-$y$) plane by regulating the tilt ($\theta_x$, $\theta_y$) of the piezo-controlled mirror, which is placed at the conjugate image plane of the back focal plane (BFP) of the objective, following Eq.~\ref{eq:back-focal}. 
        CON, OBJ, DM, and CCD in the illumination beam path (cyan lines) denote condenser, objective, dichroic mirror, which reflects the trapping laser and transmits the illumination, and a CCD camera, respectively. 
        The bottom left inset illustrates the tilt angles $\theta_x$ and $\theta_y$ around the $x$- and $y$-axis of the mirror mount. 
        The top inset shows the resultant displacement of the trap (orange gradient) at the focal plane.}%
    \label{fig:diffusing-trap}
\end{figure*}

\section*{Simulation of probe particle dynamics in an MV fluid}
We simulated the probe particle dynamics in an MV fluid to validate our model for characterizing real single-relaxation VE materials.
The numerical simulation was performed in two steps: first, we computed the position of the particle in the HW using Green's function, as $x_{\mathrm{HBBP}}(t_n) = x_{\mathrm{HBBP}}(t_{n-1}) e^{-\Delta t/\tau_k} + \sqrt{B(\Delta t)}\,R_{\mathrm{HBBP}}(t_n)$, and second, obtained the HW center position from a simple diffusion process as $x_{\mathrm{HW}}(t_n) = x_{\mathrm{HW}}(t_{n-1}) + \sqrt{2D_{\mathrm{HW}}\Delta t}\,R_{\mathrm{HW}}(t_n)$, where both $R_{\mathrm{HBBP}}(t_n)$ and $R_{\mathrm{HW}}(t_n)$ are Gaussian random numbers with zero mean and unit standard deviation.
By updating the initial HBBP position at every step ($n$), we obtained the particle position as $x(t_n) = x_{\mathrm{HBBP}}(t_n) + x_{\mathrm{HW}}(t_n)$~\cite{khanRandomWalksColloidal2014, khanTrajectoriesProbeSpheres2014}.
The resulting $\left\langle \Delta r^2 (\tau)\right\rangle$ and creep compliance $J(\tau)$ were computed from the simulated trajectories, as shown in Fig.~\ref{fig:model-mv}d. 
At short time-lags, the probe particle exhibits diffusive dynamics and experiences harmonic confinement at the equilibration time $\tau_k$, leading to a plateau in the MSD.
As the system relaxes at $\tau > \lambda (= \tau_k (D_{\mathrm{HBBP}} / D_{\mathrm{HW}}))$ due to the slow diffusion of the HW, the probe particle dynamics become diffusive again with a smaller diffusion coefficient corresponding to $D_{\mathrm{HW}}$. 
These results show excellent agreement with the experimentally observed microrheological response of single-relaxation VE fluids (Fig.~\ref{fig:msd-real}b).

\begin{figure*}[t]
  \centering
  \includegraphics[width=\linewidth]{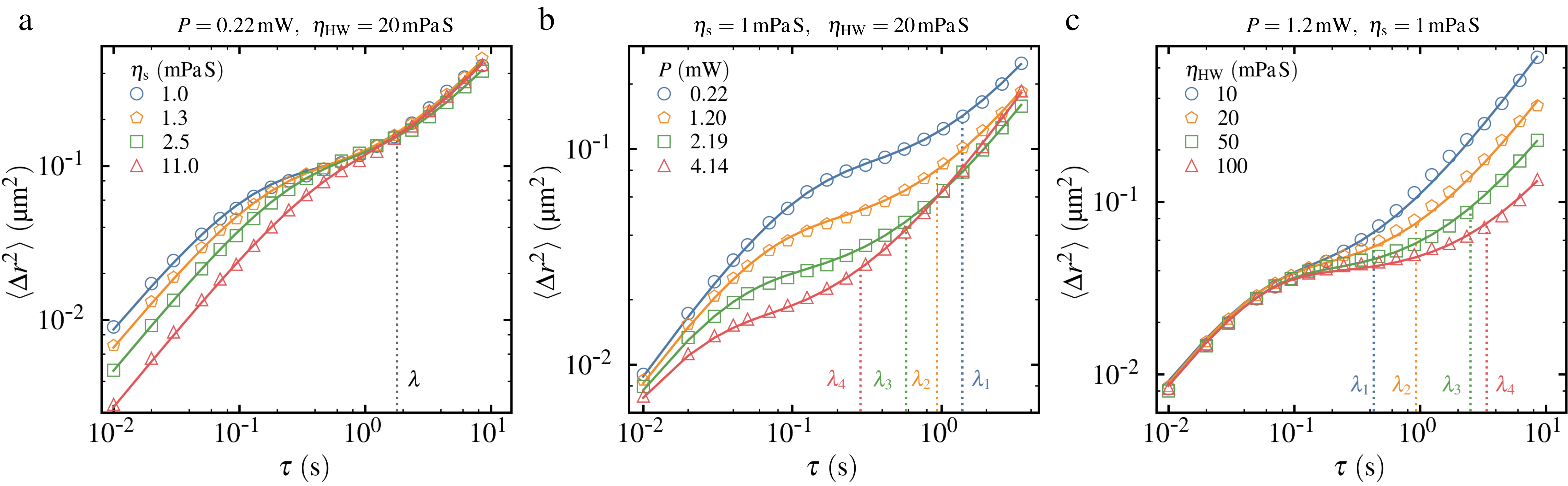}%
  \caption{Systematic experimental tuning of the VE parameters in the microrheological response of a configurable single-relaxation VE fluid,
     exhibited by the MSD of a \SI{1.98}{\micro\meter} diameter PS probe particle in a diffusive optical trap. The experimental data are shown with open symbols, whereas the solid lines of the same color represent the corresponding theoretical predictions of our model (Eq.~\ref{eq:msd-pbp}).
    (a) Systematic variation of the high-frequency viscosity $\eta_{\mathrm{s}}$ of the emergent VE environment is achieved using viscous media of different viscosities, keeping the trap stiffness $k = 6\pi a G_\mathrm{p}$ ($\propto$ laser power $P$) and the viscosity of the diffusing trap $\eta_{\mathrm{HW}}$ unchanged.
    (b) The plateau modulus $G_\mathrm{p}$ is regulated by varying the laser power $P$ while keeping $\eta_{\mathrm{s}}$ and $\eta_{\mathrm{HW}}$ unchanged. 
        An increase in $G_\mathrm{p}$ diminishes the MSD plateau value and shifts both $\tau_k = \eta_{\mathrm{s}}/G_{\mathrm{p}}$ and $\lambda = \eta_{\mathrm{HW}}/G_{\mathrm{p}}$ to shorter time-lags.
    (c) Independent variation of $\lambda$ is achieved by varying $\gamma_{\mathrm{HW}}$, i.e., $\eta_{\mathrm{HW}}$, while keeping $G_\mathrm{p}$ and $\eta_\mathrm{s}$ unchanged.}% 
  \label{fig:msd-configurable-mv}
\end{figure*}

\begin{figure}[t]
  \centering
  \includegraphics[width=62mm]{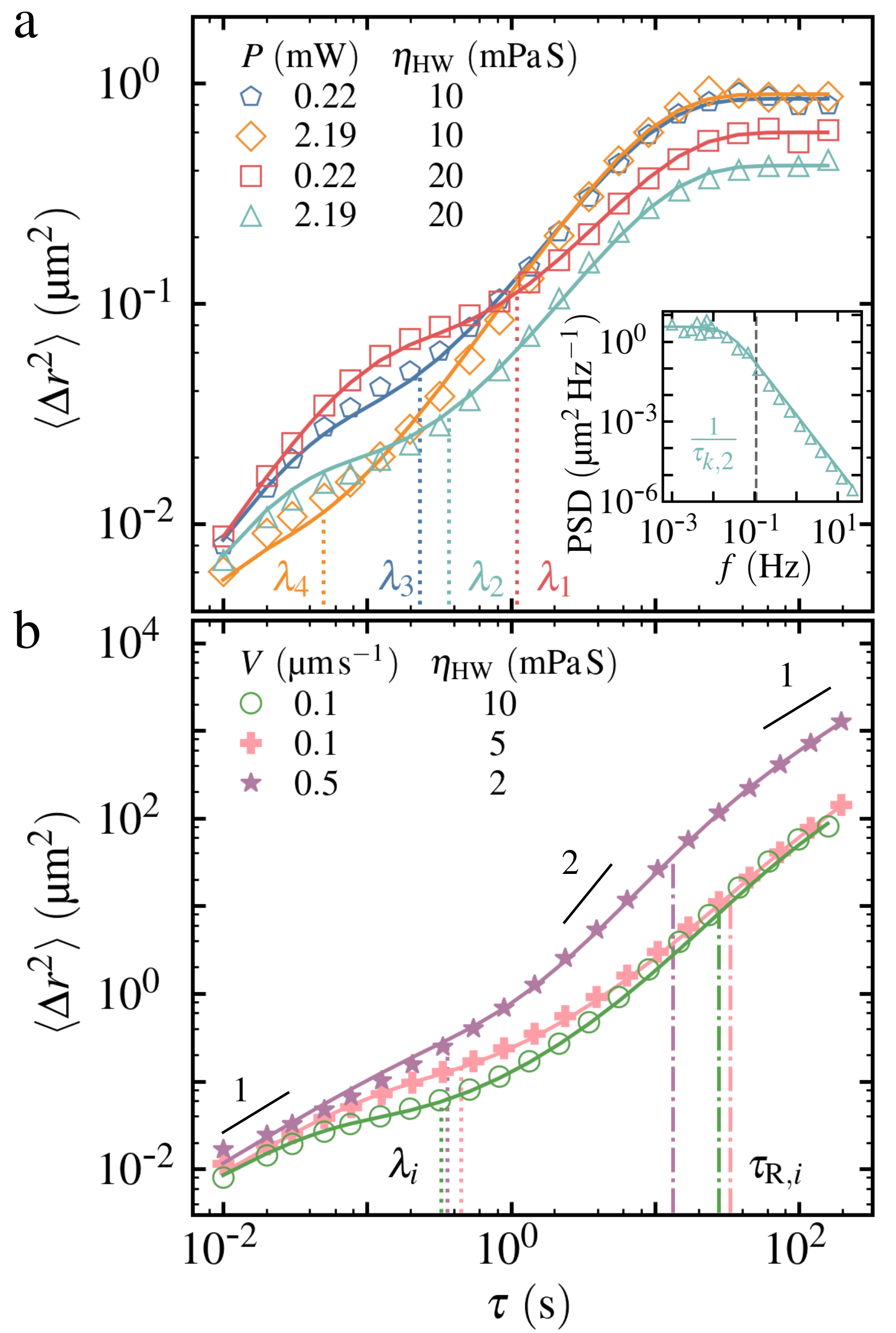}% %new
  \caption{Realization of double-relaxation VE media and active VE environments employing a dynamic trap scheme. 
    The MSDs of a \SI{1.98}{\micro\meter} diameter PS probe particle in a dynamic trap with correlated and persistent translational noise, exhibiting microrheological response of configurable double-relaxation and active VE environments. 
    The experimental data and simulation results are shown with open and filled symbols, respectively. 
    (a) Microrheological response of double-relaxation VE environments is realized by translating the trap with exponentially correlated noise, providing equilibration timescales $\tau_{k,1}$ and $\tau_{k,2}$. The power spectral density of a trap trajectory is shown in the inset.
    (b) Probe particle dynamics in active VE baths is realized by moving the trap along an ABP trajectory with persistent translational noise.
    The solid lines in (a) and (b) represent the theoretical predictions, following Eqs.~\ref{eq:msd-pbp-double} and~\ref{eq:msd-pbp-active}, respectively, corresponding to the experimentally observed MSDs of the same color.}
  \label{fig:msd-active-and-ou}
\end{figure}

\section*{Experimental realization of configurable VE environments}
We trapped a dielectric PS particle ($2a = \SI{1.98}{\micro\meter}$) in an HW created by a conventional optical tweezers setup with a linearly polarized solid-state continuous wave Nd:YAG laser at $\lambda=\SI{1064}{\nano\meter}$ operating in $\mathrm{TEM}_{00}$ Gaussian mode, as shown in Fig.~\ref{fig:diffusing-trap}a (model: Laser Quantum Opus \SI{5}{\watt}), focused through a 60$\times$ oil-immersion microscope objective with NA 1.40 (model: Nikon CFI Plan APO Lambda)~\cite{Khan2014, halderOpticalMicromanipulationSoft2024a}.
The trapped particle underwent confined Brownian motion in the HW centered at the laser focal point, with the measured MSD and position distribution corroborating the HBBP dynamics and manifesting the microrheological response of a Voigt solid (Fig.~\ref{fig:diffusing-trap}b).
Our setup incorporates a piezo-controlled steering mirror placed at the conjugate image plane of the back focal plane of the microscope objective. 
We implemented this using an afocal lens pair $L_3$ and $L_4$ (Fig.~\ref{fig:diffusing-trap}c), which ensured that the angular displacement of the mirror translated to lateral displacement of the optical trap in the focal plane~\cite{fallmanDesignFullySteerable1997, Khan2014, halderOpticalMicromanipulationSoft2024a}.
This configuration enables any predefined $x$-$y$ movement of the optical trap (Fig.~\ref{fig:diffusing-trap}c (top inset)) by controlling the tilt $\theta_x$, $\theta_y$ of the steering mirror (Fig.~\ref{fig:diffusing-trap}c (bottom left inset)).

The piezo-mirror tilt $\Delta \theta_{\mathrm{PM}}$ produces an angular displacement $\Delta \theta_{\mathrm{BFP}}$ at the back focal plane of the objective and a corresponding lateral displacement $\Delta r$ in the focal plane, given by
\begin{equation}
  \Delta \theta_{\mathrm{BFP}} = -2\frac{f_3}{f_4} \Delta \theta_{\mathrm{PM}},\qquad
  \Delta r = f_{\mathrm{EFL}} \Delta \theta_{\mathrm{BFP}},
  \label{eq:back-focal}
\end{equation}
\noindent where $f_3$ and $f_4$ are the focal lengths of lenses $L_3$ and $L_4$, respectively, and $f_{\mathrm{EFL}}$ is the effective focal length of the objective.
We generated an angular displacement time series $(\Delta \theta_x(t), \Delta \theta_y(t))$ for the steering mirror, corresponding to the required diffusion profile $(\Delta x(t), \Delta y(t))$ of the HW, and fed it to the piezo-controller. This realizes the appropriate HW dynamics for a desired relaxation behavior of the configurable VE environment.

To examine the precise and reliable movement of the HW, we tightly confined a particle using high laser power, reducing its thermal fluctuations within the trap. In this setting, the particle position closely follows the HW center. We moved the trap along a predefined Brownian trajectory by feeding the corresponding angular displacement time series to the piezo-controller and recorded the particle dynamics using a CCD camera.
The particle position was tracked using the TrackMate plugin in ImageJ~\cite{ershovTrackMate7Integrating2022}, and the obtained trajectory closely followed the programmed trap trajectory (Video~\href{https://drive.google.com/file/d/1bHBim8TPTKbDQk-7jMMZl7-r2xqlYVs7/view?usp=share_link}{S2a}).

It is important to note that the implemented dynamics of the optical trap is slower than the timescale associated with the large long-time viscosity of the emergent VE environment. Therefore, the quasistatically driven motion of the HW ensures that the probe particle remains equilibrated within the HW at all times.

The MV model describes the VE response of single-relaxation complex fluids using three parameters: high-frequency viscosity $\eta_{\mathrm{s}}$, plateau modulus $G_{\mathrm{p}}$, and relaxation time $\lambda$.
Here, we employed a dynamic optical trap to directly regulate these three parameters in the response of a trapped particle, constructing a configurable VE environment where $\eta_{\mathrm{s}}$, $G_{\mathrm{p}}$, and $\lambda$ can be tuned systematically.
Furthermore, our scheme can generate probe particle dynamics in VE media with varied and intriguing relaxation properties, some of which are difficult to realize with real materials. We demonstrated this by showing the microrheological response of configurable VE environments with double and active relaxation.

For the single-relaxation MV case, we chose the experimental controls, namely, the solvent viscosity $\eta_{\mathrm{s}}$, the stiffness of the optical trap $k$, which is governed by the power of the trapping laser $P$, and the diffusion coefficient $D_{\mathrm{HW}}$ of the trap, which are directly connected to the properties of the emergent MV fluid. This correspondence for an MV fluid is listed in Table~\ref{correspondence}. By separately varying these three experimental inputs, we generated probe particle dynamics representing the microrheological response of the corresponding MV fluid with independently varying VE parameters $\eta_{\mathrm{s}}$, $G_{\mathrm{p}}$, and $\lambda$, as demonstrated in Fig.~\ref{fig:msd-configurable-mv}.

\begin{table}[h!]
    \centering
    \caption{MV fluid VE properties, corresponding HBBP with long-time diffusion model parameters, and the controls in experimental realization of the model.}
    \label{tab:ve-parameters}
    \footnotesize
    \begin{tabular*}{80mm}{@{\extracolsep{\fill}}ccc}
        \toprule
        MV fluid & Model & Experimental \\
        VE properties & parameters & controls \\
        \midrule
        High-frequency viscosity: $\eta_{\mathrm{s}}$ & $\gamma_{\mathrm{s}} = 6\pi a \eta_{\mathrm{s}}$ & solvent viscosity \\
        Plateau modulus: $G_{\mathrm{p}}$ & $k = 6\pi a G_{\mathrm{p}}$ & $k \propto$ laser power $P$ \\
        Long-time viscosity: $\eta$ & $\gamma_{\mathrm{HW}} = 6\pi a \eta$ & trap dynamics, $D_{\mathrm{HW}}$   \\
        Relaxation time: $\lambda$ & $\lambda = \gamma_{\mathrm{HW}}/ 6\pi a G_{\mathrm{p}}$ & $P$ and $D_{\mathrm{HW}}$ \\
        \bottomrule
    \end{tabular*}
    \label{correspondence}
\end{table}

The experimentally captured probe particle MSDs are time-averaged over the entire duration of the tracked trajectories. To ensure adequate sampling, we captured long trajectories, at least 100$\times$ longer than the longest timescale of the system at a frame rate of $\sim$ \SI{1000}{\Hz} for the real MV fluid measurements and $\sim$ \SI{100}{\Hz} for the dynamic trap experiments. All experimentally observed MSDs are consistent over multiple probe particles and are shown here for time-lags up to $\sim$ 10\% of the trajectory length. The extracted VE parameters agree with the theoretical predictions within 5\% standard error.

\subsection*{Tuning high-frequency viscosity $\eta_{\mathrm{s}}$}
The high-frequency viscosity $\eta_{\mathrm{s}}$ of a VE medium governs the probe particle dynamics at short time-lags, before elastic confinement sets in and the MSD reaches a plateau at the characteristic time $\tau_k = \eta_{\mathrm{s}}/G_{\mathrm{p}}$.
Moreover, $\eta_{\mathrm{s}}$ directly regulates $\tau_k$, as they are linearly related at fixed $G_{\mathrm{p}}$.
We tuned $\eta_{\mathrm{s}}$ by choosing an appropriate viscous medium, using aqueous dilutions of glycerol with concentrations from 10 to \SI{60}{wt\%} at \SI{22}{\degreeCelsius} to vary $\eta_{\mathrm{s}}$ from 1.3 to \SI{11}{\milli\pascal\second}, well beyond pure water (\SI{1}{\milli\pascal\second})~\cite{segurViscosityGlycerolIts1951}.
With increasing $\eta_{\mathrm{s}}$, the diffusivity of the probe particle decreases and the MSDs shift downward. Hence, the plateau in the MSD is reached at longer time-lags, resulting in a longer $\tau_k$ when $G_{\mathrm{p}}$ and $\lambda$ remain unchanged. The effect of independently varying $\eta_{\mathrm{s}}$ on the MSD of a probe particle is demonstrated in Fig.~\ref{fig:msd-configurable-mv}a (Video~\href{https://drive.google.com/file/d/1bHBim8TPTKbDQk-7jMMZl7-r2xqlYVs7/view?usp=share_link}{S2b}).

\subsection*{Tuning plateau modulus $G_{\mathrm{p}}$}
The plateau modulus $G_{\mathrm{p}}$, determined from the plateau value of the MSD, corresponds directly to the elastic modulus of a Voigt solid.
In our configurable VE environment, $G_{\mathrm{p}}$ is directly controlled by the laser power $P$, as the trap stiffness $k$ varies proportionally with $P$, following $G_{\mathrm{p}} = k/6\pi a$.
Moreover, an increase in $G_{\mathrm{p}}$ shifts both the crossover time $\tau_k = \eta_{\mathrm{s}}/G_{\mathrm{p}}$ and the relaxation time $\lambda = \gamma_{\mathrm{HW}}/ 6 \pi a G_{\mathrm{p}} = \eta_{\mathrm{HW}}/G_{\mathrm{p}}$ to shorter time-lags, reducing and expanding the high- and low-frequency diffusive behaviors, respectively.
This coupled response of $\tau_k$ and $\lambda$ through $G_{\mathrm{p}}$ is a hallmark property of real single-relaxation VE materials, where increasing elasticity shortens both timescales.
The independent and systematic tuning of the plateau modulus $G_{\mathrm{p}}$ through the laser power $P$ is shown in Fig.~\ref{fig:msd-configurable-mv}b (Video~\href{https://drive.google.com/file/d/1bHBim8TPTKbDQk-7jMMZl7-r2xqlYVs7/view?usp=share_link}{S2c}).

\subsection*{Tuning relaxation time $\lambda$}
The relaxation time $\lambda$ defines the timescale at which the residual stress equilibrates and the elastic response of the VE medium decays, beyond which the medium behaves like a viscous fluid. 
In our configurable VE environment, $\lambda$ can be independently tuned by varying the viscosity $\eta_{\mathrm{HW}}$ associated with the imposed diffusive motion of the optical trap, keeping $G_{\mathrm{p}}$ and $\eta_{\mathrm{s}}$ constant.
The relaxation time $\lambda$ and the low-frequency viscosity $\eta = \gamma_{\mathrm{HW}} / 6 \pi a = \eta_{\mathrm{HW}}$ are coupled through the plateau modulus $G_{\mathrm{p}}$ as $\lambda = \gamma_{\mathrm{HW}}/6 \pi a G_{\mathrm{p}} = \eta_{\mathrm{HW}}/G_{\mathrm{p}}$.
At fixed $G_{\mathrm{p}}$, i.e., at a constant laser power, systematic regulation of the relaxation time $\lambda$ and, consequently, the low-frequency viscosity was realized by choosing an appropriate $\eta_{\mathrm{HW}}$ for the trap movement.
With increasing $\lambda$, the system relaxes more slowly and the long-time diffusive behavior in the MSD shifts to longer time-lags. 
The independent variation of $\lambda$ by regulating $\eta_{\mathrm{HW}}$ is illustrated in Fig.~\ref{fig:msd-configurable-mv}c (Video~\href{https://drive.google.com/file/d/1bHBim8TPTKbDQk-7jMMZl7-r2xqlYVs7/view?usp=share_link}{S2d}).
Such regulation requires a separation of timescales, $\tau_k = \gamma_{\mathrm{s}}/k \ll \lambda$, ensuring that the probe equilibrates within the HW faster than the trap motion.

We now extend our scheme to double-relaxation and active VE responses by modifying the trap motion with an appropriate translational noise profile.

\section*{Double-relaxation VE media}
Our scheme is extendable to realize the microrheological response of VE media with double relaxation. 
Here, we demonstrate this by showing the MSD of a probe particle in a diffusing optical trap that follows an HBBP trajectory (Video~\href{https://drive.google.com/file/d/1ggX-g-2toojGDX-9-JModgYXuvxQpzOJ/view?usp=share_link}{S3a}). 
Therefore, the trajectory in this case is generated by the same Langevin equation that describes the HBBP in the single-relaxation case, with $\tau_k$ replaced by $\tau_{k,2}$. 
The resulting trajectory is exponentially correlated, with a Lorentzian power spectral density shown as an inset in Fig.~\ref{fig:msd-active-and-ou}a.
This yields a longer equilibration timescale $\tau_{k,2}$, corresponding to the correlation time of the imposed noise, in addition to the shorter equilibration time $\tau_{k,1}$ associated with the stiffness of the optical trap.
Thereby, both equilibration timescales, and hence the two plateaus in the MSD, can be independently configured by regulating the trap stiffness through the laser power $P$ and the correlation time of the imposed noise in the trap dynamics.
Fig.~\ref{fig:msd-active-and-ou}a  shows the experimentally measured MSD of a \SI{1.98}{\micro\meter} PS particle (open symbols) displaying two distinct plateaus separated by an intermediate relaxation regime, consistent with the hierarchical VE response predicted by Eq.~\ref{eq:msd-pbp-double} (solid line).
This signifies that multiple plateaus in the MSD can be systematically configured using an appropriate noise profile with multiple correlation timescales.
Therefore, our results demonstrate that this dynamic trap framework extends beyond replicating the microrheological response of single-relaxation MV fluids and can emulate that of a broad class of VE media following the GMM with a relaxation spectrum~\cite{khanTrajectoriesProbeSpheres2014}.

\section*{Active VE environments}
Our approach can also generate probe particle dynamics in an active VE environment, such as an entangled network of active polymers~\cite{mizunoNonequilibriumMechanicsActive2007, eisensteckenInternalDynamicsSemiflexible2017, tejedorReptationActiveEntangled2019, winklerPhysicsActivePolymers2020, tejedorMolecularDynamicsSimulations2023}, by moving the trap along an ABP trajectory with persistent noise.
Fig.~\ref{fig:msd-active-and-ou}b shows the experimentally obtained MSD of a \SI{1.98}{\micro\meter} PS particle manifesting post-relaxation active dynamics (Video~\href{https://drive.google.com/file/d/1ggX-g-2toojGDX-9-JModgYXuvxQpzOJ/view?usp=share_link}{S3b}), as predicted by Eq.~\ref{eq:msd-pbp-active}.
At short time-lags $\tau < \lambda$, before relaxation, the microrheological response is similar to that of a Voigt solid, exhibiting a diffusive regime at $\tau \ll \tau_k$ with a subsequent plateau arising from elastic confinement.
At $\tau > \lambda$, the active persistent motion of the trap induces super-diffusive dynamics of the probe particle with an exponent of $\sim 2$. 
The probe dynamics become diffusive again at $\tau > \tau_{\mathrm{R}}$ as the correlation of the propulsion direction, or equivalently the persistence in the translational noise, decays.
The experimentally observed MSD (open symbols) shows excellent agreement with those from numerical simulations (filled symbols) and theoretical predictions (Eq.~\ref{eq:msd-pbp-active}, solid line).
These results demonstrate that our dynamic trap scheme provides a convenient route for studying probe particle dynamics and microrheological response of active VE environments, where the activity and VE properties can be independently and systematically regulated.

\section*{Conclusions}
In this study, we demonstrate a novel experimental scheme for realizing the microrheological response of an emergent configurable VE environment by appropriately regulating the stiffness and trajectory of the dynamic optical trap. 
While the stiffness of the optical trap regulates the plateau modulus $G_{\mathrm{p}}$ of the VE fluid, the relaxation time $\lambda$ and post-relaxation properties are governed by the trap dynamics. 
Thereby, our approach enables systematic and independent variation of the VE properties of the resultant system, overcoming the inherent coupling in real complex fluids. 
The microrheological measurements from the emergent configurable VE environment exhibit quantitative similarities with those from real complex fluids, having widely varied VE properties, validating the broad applicability of our scheme.

The key advantage of this approach lies in its experimental versatility: the systematic variation of the stiffness and dynamics of an optical trap enables the rapid exploration of VE parameter space. 
Furthermore, it provides access to VE regimes that are challenging to achieve with real complex fluids and remain mostly unaltered against variations in experimental settings, such as the temperature and surface chemistry of the probe particles.
The framework extends naturally to replicate the microrheological response of GLVE materials having multiple relaxation spectra with correlated noise in the dynamics of the trap and to active VE environments by translating the trap along an ABP trajectory.
Thus, our scheme realizes probe particle dynamics in a configurable VE fluid, providing a one-point linear microrheological response of the medium. However, a dynamic optical trap cannot generate the inter-probe spatial correlations observed in a real VE continuum probed by two-point microrheology~\cite{crockerTwoPointMicrorheologyInhomogeneous2000}.

Notably, this approach with a single dynamic optical trap cannot capture nonlinear VE responses observed in various active microrheology measurements, such as strain-stiffening, shear-thinning, strain-induced density inhomogeneities and recoil~\cite{gomez-solanoTransientDynamicsColloidal2015, khanOpticalTweezersMicrorheology2019, ginotRecoilExperimentsDetermine2022}.
However, this limitation can be overcome in future studies by using an array of weak traps that replicate the adjoining confinement in an entangled network.
Beyond passive systems, our scheme enables the study of probe particle dynamics in configurable active VE environments.
The experimental scheme reported in this study opens diverse future directions for investigating the dynamics of natural and artificial microswimmers in controlled VE environments and the VE properties of entangled active biopolymers.

\vspace{1em}
\section*{Author contributions}
All authors contributed to the conception and design of the research. 
S.H. conducted the experiments and analyzed the data. 
S.H. and M.K. interpreted the data and wrote the manuscript. 
M.K. supervised the project.

\section*{Conflict of interest}
The authors declare no competing interest.

\section*{Data availability}
All data required to reach the conclusions of this study are presented in the manuscript.

\section*{Acknowledgements}
Research funding from SERB (now ANRF), Govt.\ of India (Grant No. CRG/2020/002723) is gratefully acknowledged. M.K. acknowledges funding from IIT Kanpur through an initiation grant (grant no. IITK-PHY-2017081).

\section*{References}
\bibliographystyle{rsc} % new
\bibliography{references.bib}

@article{andablo-reyesMicrorheologyRotationalDiffusion2005,
  title = {Microrheology from {{Rotational Diffusion}} of {{Colloidal Particles}}},
  author = {{Andablo-Reyes}, Efr{\'e}n and {D{\'i}az-Leyva}, Pedro and {Arauz-Lara}, Jos{\'e} Luis},
  year = 2005,
  month = mar,
  journal = {Phys. Rev. Lett.},
  volume = {94},
  number = {10},
  pages = {106001},
  publisher = {American Physical Society (APS)},
  issn = {0031-9007, 1079-7114},
  doi = {10.1103/PhysRevLett.94.106001},
  urldate = {2023-09-01},
  copyright = {http://link.aps.org/licenses/aps-default-license},
  langid = {english}
}

@article{bellourBrownianMotionParticles2002,
  title = {Brownian Motion of Particles Embedded in a Solution of Giant Micelles},
  author = {Bellour, M. and Skouri, M. and Munch, J.-P. and H{\'e}braud, P.},
  year = 2002,
  month = jul,
  journal = {Eur. Phys. J. E},
  volume = {8},
  number = {4},
  pages = {431--436},
  issn = {1292-8941, 1292-895X},
  doi = {10.1140/epje/i2002-10026-0},
  urldate = {2025-06-07},
  copyright = {http://www.springer.com/tdm},
  langid = {english}
}

@article{cardinauxMicrorheologyGiantmicelleSolutions2002,
  title = {Microrheology of Giant-Micelle Solutions},
  author = {Cardinaux, F and Cipelletti, L and Scheffold, F and Schurtenberger, P},
  year = 2002,
  month = mar,
  journal = {Europhys. Lett.},
  volume = {57},
  number = {5},
  pages = {738--744},
  issn = {0295-5075, 1286-4854},
  doi = {10.1209/epl/i2002-00525-0},
  urldate = {2025-06-07}
}

@article{catesReptationLivingPolymers1987,
  title = {Reptation of Living Polymers: Dynamics of Entangled Polymers in the Presence of Reversible Chain-Scission Reactions},
  shorttitle = {Reptation of Living Polymers},
  author = {Cates, M. E.},
  year = 1987,
  month = sep,
  journal = {Macromolecules},
  volume = {20},
  number = {9},
  pages = {2289--2296},
  issn = {0024-9297, 1520-5835},
  doi = {10.1021/ma00175a038},
  urldate = {2025-06-07},
  langid = {english}
}

@article{catesStaticsDynamicsWormlike1990,
  title = {Statics and Dynamics of Worm-like Surfactant Micelles},
  author = {Cates, M E and Candau, S J},
  year = 1990,
  month = aug,
  journal = {J. Phys.: Condens. Matter},
  volume = {2},
  number = {33},
  pages = {6869--6892},
  issn = {0953-8984, 1361-648X},
  doi = {10.1088/0953-8984/2/33/001},
  urldate = {2025-06-05}
}

@article{celliHelicobacterPyloriMoves2009,
  title = {{\emph{Helicobacter Pylori}} Moves through Mucus by Reducing Mucin Viscoelasticity},
  author = {Celli, Jonathan P. and Turner, Bradley S. and Afdhal, Nezam H. and Keates, Sarah and Ghiran, Ionita and Kelly, Ciaran P. and Ewoldt, Randy H. and McKinley, Gareth H. and So, Peter and Erramilli, Shyamsunder and Bansil, Rama},
  year = 2009,
  month = aug,
  journal = {Proc. Natl. Acad. Sci. U.S.A.},
  volume = {106},
  number = {34},
  pages = {14321--14326},
  issn = {0027-8424, 1091-6490},
  doi = {10.1073/pnas.0903438106},
  urldate = {2025-09-20},
  langid = {english}
}

@article{chandrasekharStochasticProblemsPhysics1943,
  title = {Stochastic {{Problems}} in {{Physics}} and {{Astronomy}}},
  author = {Chandrasekhar, S.},
  year = 1943,
  month = jan,
  journal = {Rev. Mod. Phys.},
  volume = {15},
  number = {1},
  pages = {1--89},
  issn = {0034-6861},
  doi = {10.1103/RevModPhys.15.1},
  urldate = {2023-09-01},
  langid = {english}
}

@article{chengRotationalDiffusionMicrorheology2003,
  title = {Rotational {{Diffusion Microrheology}}},
  author = {Cheng, Z. and Mason, T. G.},
  year = 2003,
  month = jan,
  journal = {Phys. Rev. Lett.},
  volume = {90},
  number = {1},
  pages = {018304},
  issn = {0031-9007, 1079-7114},
  doi = {10.1103/PhysRevLett.90.018304},
  urldate = {2023-09-01},
  copyright = {http://link.aps.org/licenses/aps-default-license},
  langid = {english}
}

@article{crockerTwoPointMicrorheologyInhomogeneous2000,
  title = {Two-{{Point Microrheology}} of {{Inhomogeneous Soft Materials}}},
  author = {Crocker, John C. and Valentine, M. T. and Weeks, Eric R. and Gisler, T. and Kaplan, P. D. and Yodh, A. G. and Weitz, D. A.},
  year = 2000,
  month = jul,
  journal = {Phys. Rev. Lett.},
  volume = {85},
  number = {4},
  pages = {888--891},
  issn = {0031-9007, 1079-7114},
  doi = {10.1103/PhysRevLett.85.888},
  urldate = {2025-08-21},
  copyright = {http://link.aps.org/licenses/aps-default-license},
  langid = {english}
}

@article{dasguptaMicrorheologyPolyethyleneOxide2002,
  title = {Microrheology of Polyethylene Oxide Using Diffusing Wave Spectroscopy and Single Scattering},
  author = {Dasgupta, Bivash R. and Tee, Shang-You and Crocker, John C. and Frisken, B. J. and Weitz, D. A.},
  year = 2002,
  month = may,
  journal = {Phys. Rev. E},
  volume = {65},
  number = {5},
  pages = {051505},
  issn = {1063-651X, 1095-3787},
  doi = {10.1103/PhysRevE.65.051505},
  urldate = {2024-01-26},
  copyright = {http://link.aps.org/licenses/aps-default-license},
  langid = {english}
}

@book{doiTheoryPolymerDynamics2013,
  title = {The Theory of Polymer Dynamics},
  author = {Doi, Masao and Edwards, Samuel F.},
  year = 2013,
  series = {International Series of Monographs on Physics},
  edition = {Reprint},
  number = {73},
  publisher = {Clarendon Press},
  address = {Oxford},
  isbn = {978-0-19-852033-7},
  langid = {english}
}

@article{eisensteckenInternalDynamicsSemiflexible2017,
  title = {Internal Dynamics of Semiflexible Polymers with Active Noise},
  author = {Eisenstecken, Thomas and Gompper, Gerhard and Winkler, Roland G.},
  year = 2017,
  month = apr,
  journal = {The Journal of Chemical Physics},
  volume = {146},
  number = {15},
  pages = {154903},
  issn = {0021-9606, 1089-7690},
  doi = {10.1063/1.4981012},
  urldate = {2025-12-07},
  langid = {english}
}

@article{ershovTrackMate7Integrating2022,
  title = {{{TrackMate}} 7: Integrating State-of-the-Art Segmentation Algorithms into Tracking Pipelines},
  shorttitle = {{{TrackMate}} 7},
  author = {Ershov, Dmitry and Phan, Minh-Son and Pylv{\"a}n{\"a}inen, Joanna W. and Rigaud, St{\'e}phane U. and Le Blanc, Laure and {Charles-Orszag}, Arthur and Conway, James R. W. and Laine, Romain F. and Roy, Nathan H. and Bonazzi, Daria and Dum{\'e}nil, Guillaume and Jacquemet, Guillaume and Tinevez, Jean-Yves},
  year = 2022,
  month = jul,
  journal = {Nat Methods},
  volume = {19},
  number = {7},
  pages = {829--832},
  issn = {1548-7091, 1548-7105},
  doi = {10.1038/s41592-022-01507-1},
  urldate = {2024-12-12},
  langid = {english}
}

@article{fallmanDesignFullySteerable1997,
  title = {Design for Fully Steerable Dual-Trap Optical Tweezers},
  author = {F{\"a}llman, Erik and Axner, Ove},
  year = 1997,
  month = apr,
  journal = {Appl. Opt.},
  volume = {36},
  number = {10},
  pages = {2107},
  publisher = {Optica Publishing Group},
  issn = {0003-6935, 1539-4522},
  doi = {10.1364/ao.36.002107},
  urldate = {2025-07-19},
  copyright = {https://doi.org/10.1364/OA\_License\_v1\#VOR},
  langid = {english}
}

@book{ferryViscoelasticPropertiesPolymers1980,
  title = {Viscoelastic Properties of Polymers},
  author = {Ferry, John D.},
  year = 1980,
  edition = {3d ed},
  publisher = {Wiley},
  address = {New York},
  isbn = {978-0-471-04894-7},
  lccn = {TA455.P58 F4 1980}
}

@book{furstMicrorheology2020,
  title = {Microrheology},
  author = {Furst, Eric M. and Squires, Todd M.},
  year = 2020,
  edition = {First published in paperback},
  publisher = {Oxford University Press},
  address = {Oxford},
  isbn = {978-0-19-965520-5 978-0-19-886709-8},
  langid = {english}
}

@article{gardelMicrorheologyEntangledFActin2003,
  title = {Microrheology of {{Entangled F-Actin Solutions}}},
  author = {Gardel, M. L. and Valentine, M. T. and Crocker, J. C. and Bausch, A. R. and Weitz, D. A.},
  year = 2003,
  month = oct,
  journal = {Phys. Rev. Lett.},
  volume = {91},
  number = {15},
  pages = {158302},
  issn = {0031-9007, 1079-7114},
  doi = {10.1103/PhysRevLett.91.158302},
  urldate = {2025-08-21},
  copyright = {http://link.aps.org/licenses/aps-default-license},
  langid = {english}
}

@article{ginotRecoilExperimentsDetermine2022,
  title = {Recoil Experiments Determine the Eigenmodes of Viscoelastic Fluids},
  author = {Ginot, F{\'e}lix and Caspers, Juliana and Reinalter, Luis Frieder and Krishna Kumar, Karthika and Kr{\"u}ger, Matthias and Bechinger, Clemens},
  year = 2022,
  month = dec,
  journal = {New J. Phys.},
  volume = {24},
  number = {12},
  pages = {123013},
  issn = {1367-2630},
  doi = {10.1088/1367-2630/aca8c7},
  urldate = {2025-06-07}
}

@article{gittesMicroscopicViscoelasticityShear1997,
  title = {Microscopic {{Viscoelasticity}}: {{Shear Moduli}} of {{Soft Materials Determined}} from {{Thermal Fluctuations}}},
  shorttitle = {Microscopic {{Viscoelasticity}}},
  author = {Gittes, F. and Schnurr, B. and Olmsted, P. D. and MacKintosh, F. C. and Schmidt, C. F.},
  year = 1997,
  month = oct,
  journal = {Phys. Rev. Lett.},
  volume = {79},
  number = {17},
  pages = {3286--3289},
  issn = {0031-9007, 1079-7114},
  doi = {10.1103/PhysRevLett.79.3286},
  urldate = {2026-01-29},
  copyright = {http://link.aps.org/licenses/aps-default-license},
  langid = {english}
}

@article{gomez-solanoDynamicsSelfPropelledJanus2016,
  title = {Dynamics of {{Self-Propelled Janus Particles}} in {{Viscoelastic Fluids}}},
  author = {{Gomez-Solano}, Juan Ruben and Blokhuis, Alex and Bechinger, Clemens},
  year = 2016,
  month = mar,
  journal = {Phys. Rev. Lett.},
  volume = {116},
  number = {13},
  pages={138301},
  publisher = {American Physical Society (APS)},
  issn = {0031-9007, 1079-7114},
  doi = {10.1103/physrevlett.116.138301},
  urldate = {2025-07-20},
  copyright = {http://link.aps.org/licenses/aps-default-license},
  langid = {english}
}

@article{gomez-solanoTransientDynamicsColloidal2015,
  title = {Transient Dynamics of a Colloidal Particle Driven through a Viscoelastic Fluid},
  author = {{Gomez-Solano}, Juan Ruben and Bechinger, Clemens},
  year = 2015,
  month = oct,
  journal = {New J. Phys.},
  volume = {17},
  number = {10},
  pages = {103032},
  issn = {1367-2630},
  doi = {10.1088/1367-2630/17/10/103032},
  urldate = {2025-06-07}
}

@article{grimmBrownianMotionMaxwell2011,
  title = {Brownian Motion in a {{Maxwell}} Fluid},
  author = {Grimm, Matthias and Jeney, Sylvia and Franosch, Thomas},
  year = 2011,
  journal = {Soft Matter},
  volume = {7},
  number = {5},
  pages = {2076},
  issn = {1744-683X, 1744-6848},
  doi = {10.1039/c0sm00636j},
  urldate = {2025-06-07},
  langid = {english}
}

@incollection{halderOpticalMicromanipulationSoft2024a,
  title = {Optical {{Micromanipulation}} of {{Soft Materials}}: {{Applications}} in {{Devices}} and {{Technologies}}},
  shorttitle = {Optical {{Micromanipulation}} of {{Soft Materials}}},
  booktitle = {Advanced {{Structured Materials}}},
  author = {Halder, Sanatan and Chanda, Debojit and Mondal, Dibyendu and Kundu, Sandip and Khan, Manas},
  year = 2024,
  pages = {415--469},
  publisher = {Springer Nature Singapore},
  address = {Singapore},
  issn = {1869-8433, 1869-8441},
  doi = {10.1007/978-981-97-9468-3_13},
  urldate = {2025-07-16},
  copyright = {https://www.springernature.com/gp/researchers/text-and-data-mining},
  isbn = {978-981-97-9467-6 978-981-97-9468-3},
  langid = {english}
}

@article{hassanMicrorheologyWormlikeMicellar2005,
  title = {Microrheology of {{Wormlike Micellar Fluids}} from the {{Diffusion}} of {{Colloidal Probes}}},
  author = {Hassan, P. A. and Bhattacharya, K. and Kulshreshtha, S. K. and Raghavan, S. R.},
  year = 2005,
  month = may,
  journal = {J. Phys. Chem. B},
  volume = {109},
  number = {18},
  pages = {8744--8748},
  issn = {1520-6106, 1520-5207},
  doi = {10.1021/jp0442807},
  urldate = {2025-06-07},
  langid = {english}
}

@article{kernRheologicalPropertiesSemidilute1992,
  title = {Rheological Properties of Semidilute and Concentrated Aqueous Solutions of Cetyltrimethylammonium Bromide in the Presence of Potassium Bromide},
  author = {Kern, F. and Lemarechal, P. and Candau, S. J. and Cates, M. E.},
  year = 1992,
  month = feb,
  journal = {Langmuir},
  volume = {8},
  number = {2},
  pages = {437--440},
  issn = {0743-7463, 1520-5827},
  doi = {10.1021/la00038a020},
  urldate = {2025-06-07},
  langid = {english}
}

@article{khanOpticalTweezersMicrorheology2019,
  title = {Optical {{Tweezers Microrheology Maps}} the {{Dynamics}} of {{Strain-Induced Local Inhomogeneities}} in {{Entangled Polymers}}},
  author = {Khan, Manas and Regan, Kathryn and {Robertson-Anderson}, Rae M.},
  year = 2019,
  month = jul,
  journal = {Phys. Rev. Lett.},
  volume = {123},
  number = {3},
  pages = {038001},
  issn = {0031-9007, 1079-7114},
  doi = {10.1103/PhysRevLett.123.038001},
  urldate = {2025-07-30},
  langid = {english}
}

@article{khanRandomWalksColloidal2014,
  title = {Random Walks of Colloidal Probes in Viscoelastic Materials},
  author = {Khan, Manas and Mason, Thomas G.},
  year = 2014,
  month = apr,
  journal = {Phys. Rev. E},
  volume = {89},
  number = {4},
  pages = {042309},
  issn = {1539-3755, 1550-2376},
  doi = {10.1103/PhysRevE.89.042309},
  urldate = {2023-09-01},
  langid = {english}
}

@article{khanTrajectoriesProbeSpheres2014,
  title = {Trajectories of Probe Spheres in Generalized Linear Viscoelastic Complex Fluids},
  author = {Khan, Manas and Mason, Thomas G.},
  year = 2014,
  month = sep,
  journal = {Soft Matter},
  volume = {10},
  number = {45},
  pages = {9073--9081},
  issn = {1744-683X, 1744-6848},
  doi = {10.1039/C4SM01795A},
  urldate = {2023-09-01},
  langid = {english}
}

@article{lauMicrorheologyStressFluctuations2003,
  title = {Microrheology, {{Stress Fluctuations}}, and {{Active Behavior}} of {{Living Cells}}},
  author = {Lau, A. W. C. and Hoffman, B. D. and Davies, A. and Crocker, J. C. and Lubensky, T. C.},
  year = 2003,
  month = nov,
  journal = {Phys. Rev. Lett.},
  volume = {91},
  number = {19},
  pages = {198101},
  issn = {0031-9007, 1079-7114},
  doi = {10.1103/PhysRevLett.91.198101},
  urldate = {2025-08-21},
  copyright = {http://link.aps.org/licenses/aps-default-license},
  langid = {english}
}

@article{luProbeSizeEffects2002,
  title = {Probe Size Effects on the Microrheology of Associating Polymer Solutions},
  author = {Lu, Qiang and Solomon, Michael J.},
  year = 2002,
  month = dec,
  journal = {Phys. Rev. E},
  volume = {66},
  number = {6},
  pages = {061504},
  issn = {1063-651X, 1095-3787},
  doi = {10.1103/PhysRevE.66.061504},
  urldate = {2026-01-16},
  copyright = {http://link.aps.org/licenses/aps-default-license},
  langid = {english}
}

@article{masonDiffusingwavespectroscopyMeasurementsViscoelasticity1997,
  title = {Diffusing-Wave-Spectroscopy Measurements of Viscoelasticity of Complex Fluids},
  author = {Mason, T. G. and Gang, Hu and Weitz, D. A.},
  year = 1997,
  month = jan,
  journal = {J. Opt. Soc. Am. A},
  volume = {14},
  number = {1},
  pages = {139},
  issn = {1084-7529, 1520-8532},
  doi = {10.1364/JOSAA.14.000139},
  urldate = {2025-07-30},
  copyright = {https://doi.org/10.1364/OA\_License\_v1\#VOR},
  langid = {english}
}

@article{masonEstimatingViscoelasticModuli2000,
  title = {Estimating the Viscoelastic Moduli of Complex Fluids Using the Generalized {{Stokes--Einstein}} Equation},
  author = {Mason, Thomas G},
  year = 2000,
  journal = {Rheologica acta},
  volume = {39},
  number = {4},
  pages = {371--378}
}

@article{masonOpticalMeasurementsFrequencyDependent1995,
  title = {Optical {{Measurements}} of {{Frequency-Dependent Linear Viscoelastic Moduli}} of {{Complex Fluids}}},
  author = {Mason, T. G. and Weitz, D. A.},
  year = 1995,
  month = feb,
  journal = {Phys. Rev. Lett.},
  volume = {74},
  number = {7},
  pages = {1250--1253},
  issn = {0031-9007, 1079-7114},
  doi = {10.1103/PhysRevLett.74.1250},
  urldate = {2024-01-26},
  langid = {english}
}

@article{masonParticleTrackingMicrorheology1997,
  title = {Particle {{Tracking Microrheology}} of {{Complex Fluids}}},
  author = {Mason, T. G. and Ganesan, K. and Van Zanten, J. H. and Wirtz, D. and Kuo, S. C.},
  year = 1997,
  month = oct,
  journal = {Phys. Rev. Lett.},
  volume = {79},
  number = {17},
  pages = {3282--3285},
  issn = {0031-9007, 1079-7114},
  doi = {10.1103/PhysRevLett.79.3282},
  urldate = {2025-06-07},
  copyright = {http://link.aps.org/licenses/aps-default-license},
  langid = {english}
}

@article{masonRheologyComplexFluids1996,
  title = {Rheology of Complex Fluids Measured by Dynamic Light Scattering},
  author = {Mason, T.G. and Gang, Hu and Weitz, D.A.},
  year = 1996,
  month = sep,
  journal = {Journal of Molecular Structure},
  volume = {383},
  number = {1-3},
  pages = {81--90},
  issn = {00222860},
  doi = {10.1016/S0022-2860(96)09272-1},
  urldate = {2025-06-07},
  langid = {english}
}

@article{mcgrathMechanicsFActinMicroenvironments2000,
  title = {The {{Mechanics}} of {{F-Actin Microenvironments Depend}} on the {{Chemistry}} of {{Probing Surfaces}}},
  author = {McGrath, James L. and Hartwig, John H. and Kuo, Scot C.},
  year = 2000,
  month = dec,
  journal = {Biophysical Journal},
  volume = {79},
  number = {6},
  pages = {3258--3266},
  issn = {00063495},
  doi = {10.1016/S0006-3495(00)76558-1},
  urldate = {2026-01-16},
  copyright = {https://www.elsevier.com/tdm/userlicense/1.0/},
  langid = {english}
}

@article{mizunoNonequilibriumMechanicsActive2007,
  title = {Nonequilibrium {{Mechanics}} of {{Active Cytoskeletal Networks}}},
  author = {Mizuno, Daisuke and Tardin, Catherine and Schmidt, C. F. and MacKintosh, F. C.},
  year = 2007,
  month = jan,
  journal = {Science},
  volume = {315},
  number = {5810},
  pages = {370--373},
  issn = {0036-8075, 1095-9203},
  doi = {10.1126/science.1134404},
  urldate = {2026-01-16},
  langid = {english}
}

@article{pattesonRunningTumblingColi2015,
  title = {Running and Tumbling with {{E}}. Coli in Polymeric Solutions},
  author = {Patteson, A. E. and Gopinath, A. and Goulian, M. and Arratia, P. E.},
  year = 2015,
  month = oct,
  journal = {Sci Rep},
  volume = {5},
  number = {1},
  pages={15761},
  publisher = {{Springer Science and Business Media LLC}},
  issn = {2045-2322},
  doi = {10.1038/srep15761},
  urldate = {2025-07-18},
  copyright = {https://creativecommons.org/licenses/by/4.0},
  langid = {english}
}

@article{raikherBrownianMotionViscoelastic2013,
  title = {Brownian Motion in a Viscoelastic Medium Modelled by a {{Jeffreys}} Fluid},
  author = {Raikher, Yuriy L. and Rusakov, Victor V. and Perzynski, R{\'e}gine},
  year = 2013,
  journal = {Soft Matter},
  volume = {9},
  number = {45},
  pages = {10857},
  issn = {1744-683X, 1744-6848},
  doi = {10.1039/c3sm51956b},
  urldate = {2025-06-07},
  langid = {english}
}

@article{raikherTheoryBrownianMotion2010,
  title = {Theory of {{Brownian}} Motion in a {{Jeffreys}} Fluid},
  author = {Raikher, {\relax Yu}. L. and Rusakov, V. V.},
  year = 2010,
  month = nov,
  journal = {J. Exp. Theor. Phys.},
  volume = {111},
  number = {5},
  pages = {883--889},
  issn = {1063-7761, 1090-6509},
  doi = {10.1134/S1063776110110191},
  urldate = {2025-06-07},
  copyright = {http://www.springer.com/tdm},
  langid = {english}
}

@article{rusakovBrownianMotionFluids2015,
  title = {Brownian {{Motion}} in the {{Fluids}} with {{Complex Rheology}}},
  author = {Rusakov, V. V. and Raikher, {\relax Yu}. L. and Perzynski, R.},
  editor = {Nepomnyashchy, A. A.},
  year = 2015,
  journal = {Math. Model. Nat. Phenom.},
  volume = {10},
  number = {4},
  pages = {1--43},
  issn = {0973-5348, 1760-6101},
  doi = {10.1051/mmnp/201510401},
  urldate = {2025-06-07}
}

@article{segurViscosityGlycerolIts1951,
  title = {Viscosity of {{Glycerol}} and {{Its Aqueous Solutions}}},
  author = {Segur, J. B. and Oberstar, Helen E.},
  year = 1951,
  month = sep,
  journal = {Ind. Eng. Chem.},
  volume = {43},
  number = {9},
  pages = {2117--2120},
  issn = {0019-7866, 1541-5724},
  doi = {10.1021/ie50501a040},
  urldate = {2025-06-07},
  langid = {english}
}

@article{shabaniverkiCharacterizingGelatinHydrogel2017,
  title = {Characterizing Gelatin Hydrogel Viscoelasticity with Diffusing Colloidal Probe Microscopy},
  author = {Shabaniverki, Soheila and Ju{\'a}rez, Jaime J.},
  year = 2017,
  month = jul,
  journal = {Journal of Colloid and Interface Science},
  volume = {497},
  pages = {73--82},
  issn = {00219797},
  doi = {10.1016/j.jcis.2017.02.057},
  urldate = {2025-06-07},
  langid = {english}
}

@article{shikataNonlinearViscoelasticBehavior1988,
  title = {Nonlinear Viscoelastic Behavior of Aqueous Detergent Solutions},
  author = {Shikata, Toshiyuki and Hirata, Hirotaka and Takatori, Eiichi and Osaki, Kunihiro},
  year = 1988,
  month = jan,
  journal = {Journal of Non-Newtonian Fluid Mechanics},
  volume = {28},
  number = {2},
  pages = {171--182},
  issn = {03770257},
  doi = {10.1016/0377-0257(88)85038-9},
  urldate = {2025-06-05},
  copyright = {https://www.elsevier.com/tdm/userlicense/1.0/},
  langid = {english}
}

@article{squiresFluidMechanicsMicrorheology2010,
  title = {Fluid {{Mechanics}} of {{Microrheology}}},
  author = {Squires, Todd M. and Mason, Thomas G.},
  year = 2010,
  month = jan,
  journal = {Annu. Rev. Fluid Mech.},
  volume = {42},
  number = {1},
  pages = {413--438},
  issn = {0066-4189, 1545-4479},
  doi = {10.1146/annurev-fluid-121108-145608},
  urldate = {2025-06-07},
  langid = {english}
}

@article{squiresNonlinearMicrorheologyBulk2008,
  title = {Nonlinear {{Microrheology}}: {{Bulk Stresses}} versus {{Direct Interactions}}},
  shorttitle = {Nonlinear {{Microrheology}}},
  author = {Squires, Todd M.},
  year = 2008,
  month = feb,
  journal = {Langmuir},
  volume = {24},
  number = {4},
  pages = {1147--1159},
  issn = {0743-7463, 1520-5827},
  doi = {10.1021/la7023692},
  urldate = {2025-06-07},
  langid = {english}
}

@article{tejedorMolecularDynamicsSimulations2023,
  title = {Molecular Dynamics Simulations of Active Entangled Polymers Reptating through a Passive Mesh},
  author = {Tejedor, Andr{\'e}s R. and Carracedo, Raquel and Ram{\'i}rez, Jorge},
  year = 2023,
  month = feb,
  journal = {Polymer},
  volume = {268},
  pages = {125677},
  issn = {00323861},
  doi = {10.1016/j.polymer.2023.125677},
  urldate = {2025-07-30},
  langid = {english}
}

@article{tejedorReptationActiveEntangled2019,
  title = {Reptation of {{Active Entangled Polymers}}},
  author = {Tejedor, Andr{\'e}s R. and Ram{\'i}rez, Jorge},
  year = 2019,
  month = nov,
  journal = {Macromolecules},
  volume = {52},
  number = {22},
  pages = {8788--8792},
  issn = {0024-9297, 1520-5835},
  doi = {10.1021/acs.macromol.9b01994},
  urldate = {2025-07-30},
  copyright = {https://doi.org/10.15223/policy-029},
  langid = {english}
}

@article{uhlenbeckTheoryBrownianMotion1930,
  title = {On the {{Theory}} of the {{Brownian Motion}}},
  author = {Uhlenbeck, G. E. and Ornstein, L. S.},
  year = 1930,
  month = sep,
  journal = {Phys. Rev.},
  volume = {36},
  number = {5},
  pages = {823--841},
  issn = {0031-899X},
  doi = {10.1103/PhysRev.36.823},
  urldate = {2023-09-01},
  langid = {english}
}

@article{vanzantenBrownianMotionColloidal2004,
  title = {Brownian {{Motion}} of {{Colloidal Spheres}} in {{Aqueous PEO Solutions}}},
  author = {Van Zanten, John H. and Amin, Samiul and Abdala, Ahmed A.},
  year = 2004,
  month = may,
  journal = {Macromolecules},
  volume = {37},
  number = {10},
  pages = {3874--3880},
  issn = {0024-9297, 1520-5835},
  doi = {10.1021/ma035250p},
  urldate = {2025-06-07},
  langid = {english}
}

@article{vanzantenBrownianMotionSingle2000,
  title = {Brownian Motion in a Single Relaxation Time {{Maxwell}} Fluid},
  author = {Van Zanten, John H. and Rufener, Karl P.},
  year = 2000,
  month = oct,
  journal = {Phys. Rev. E},
  volume = {62},
  number = {4},
  pages = {5389--5396},
  issn = {1063-651X, 1095-3787},
  doi = {10.1103/PhysRevE.62.5389},
  urldate = {2025-06-07},
  copyright = {http://link.aps.org/licenses/aps-default-license},
  langid = {english}
}

@article{wilhelmMagneticNanoparticlesInternal2009,
  title = {Magnetic Nanoparticles: {{Internal}} Probes and Heaters within Living Cells},
  shorttitle = {Magnetic Nanoparticles},
  author = {Wilhelm, Claire and Gazeau, Florence},
  year = 2009,
  month = apr,
  journal = {Journal of Magnetism and Magnetic Materials},
  volume = {321},
  number = {7},
  pages = {671--674},
  issn = {03048853},
  doi = {10.1016/j.jmmm.2008.11.022},
  urldate = {2025-08-09},
  copyright = {https://www.elsevier.com/tdm/userlicense/1.0/},
  langid = {english}
}

@article{wilhelmRotationalMagneticEndosome2003,
  title = {Rotational Magnetic Endosome Microrheology: {{Viscoelastic}} Architecture inside Living Cells},
  shorttitle = {Rotational Magnetic Endosome Microrheology},
  author = {Wilhelm, C. and Gazeau, F. and Bacri, J.-C.},
  year = 2003,
  month = jun,
  journal = {Phys. Rev. E},
  volume = {67},
  number = {6},
  pages = {061908},
  issn = {1063-651X, 1095-3787},
  doi = {10.1103/PhysRevE.67.061908},
  urldate = {2025-08-09},
  copyright = {http://link.aps.org/licenses/aps-default-license},
  langid = {english}
}

@article{wilkingOpticallyDrivenNonlinear2008,
  title = {Optically Driven Nonlinear Microrheology of Gelatin},
  author = {Wilking, James N. and Mason, Thomas G.},
  year = 2008,
  month = may,
  journal = {Phys. Rev. E},
  volume = {77},
  number = {5},
  pages = {055101},
  issn = {1539-3755, 1550-2376},
  doi = {10.1103/PhysRevE.77.055101},
  urldate = {2024-01-26},
  copyright = {http://link.aps.org/licenses/aps-default-license},
  langid = {english}
}

@article{winklerPhysicsActivePolymers2020,
  title = {The Physics of Active Polymers and Filaments},
  author = {Winkler, Roland G. and Gompper, Gerhard},
  year = 2020,
  month = jul,
  journal = {The Journal of Chemical Physics},
  volume = {153},
  number = {4},
  pages = {040901},
  issn = {0021-9606, 1089-7690},
  doi = {10.1063/5.0011466},
  urldate = {2025-11-05},
  langid = {english}
}

@Book{Khan2014,
  author    = {Manas Khan},
  publisher = {Scholars' Press},
  title     = {Optical Tweezers Methods and Applications in Contemporary Research},
  year      = {2014},
  isbn      = {978-3639518795},
}

@article{tenhagenNonGaussianBehaviourSelfpropelled2009,
  title = {Non-{{Gaussian}} Behaviour of a Self-Propelled Particle on a Substrate},
  author = {{Ten Hagen} and {Van Teeffelen} and {Lowen}},
  year = 2009,
  journal = {Condens. Matter Phys.},
  volume = {12},
  number = {4},
  pages = {725--738},
  issn = {1607324X},
  doi = {10.5488/CMP.12.4.725},
  urldate = {2025-08-29}
}
\end{document}